\documentclass[aps,prl,twocolumn,amsmath,amssymb,superscriptaddress,citeautoscript,floatfix]{revtex4-2}
\let\oldAA\AA
\renewcommand{\AA}{\text{\normalfont\oldAA}}

\usepackage[allow-number-unit-breaks]{siunitx}
\usepackage{graphicx}
\usepackage{dcolumn}
\usepackage{tabularx}
\usepackage{bm}
\usepackage{color}

\usepackage{mathrsfs}
\usepackage{amsfonts}
\usepackage{multirow}
\usepackage{amsmath}
\usepackage[colorlinks=true,linkcolor=blue,filecolor=cyan,citecolor=blue,urlcolor=blue]{hyperref}
\usepackage{natbib}
\setcitestyle{super}

\begin{document}

\title{Renormalizing Antiferroelectric Nanostripes in $\beta'-\rm{In}_{2}\rm{Se}_{3}$ via Optomechanics}

\author{Zihang Wu}
\affiliation{Center for Alloy Innovation and Design, State Key Laboratory for Mechanical Behavior of Materials, Xi'an Jiaotong University, Xi'an, 710049, China}

\author{Kun Liu}
\affiliation{Center for Alloy Innovation and Design, State Key Laboratory for Mechanical Behavior of Materials, Xi'an Jiaotong University, Xi'an, 710049, China}

\author{Xingchi Mu}
\affiliation{Center for Alloy Innovation and Design, State Key Laboratory for Mechanical Behavior of Materials, Xi'an Jiaotong University, Xi'an, 710049, China}

\author{Jian Zhou}\email{jianzhou@xjtu.edu.cn}
\affiliation{Center for Alloy Innovation and Design, State Key Laboratory for Mechanical Behavior of Materials, Xi'an Jiaotong University, Xi'an, 710049, China}

\begin{abstract}
Antiferroelectric (AFE) materials have received tremendous attention owing to their high energy conversion efficiency and good tunability. Recently, an exotic two-dimensional (2D) AFE material, $\beta'-\rm In_2Se_3$ monolayer that could host atomically thin AFE nanostripe domains has been experimentally synthesized and theoretically examined. In this work, we apply first-principles calculations and theoretical estimations to predict that light irradiation can control the nanostripe width of such a system. We suggest that an intermediate near-infrared light (below bandgap) could effectively harness the thermodynamic Gibbs free energy, and the AFE nanostripe width will gradually reduce. We also propose to use an above bandgap linearly polarized light to generate AFE nanostripe-specific photocurrent, providing an all-optical pump-probe setup for such AFE nanostripe width phase transitions.

\par
\centering
\textbf{TOC Graphic}

\includegraphics[width=2in]{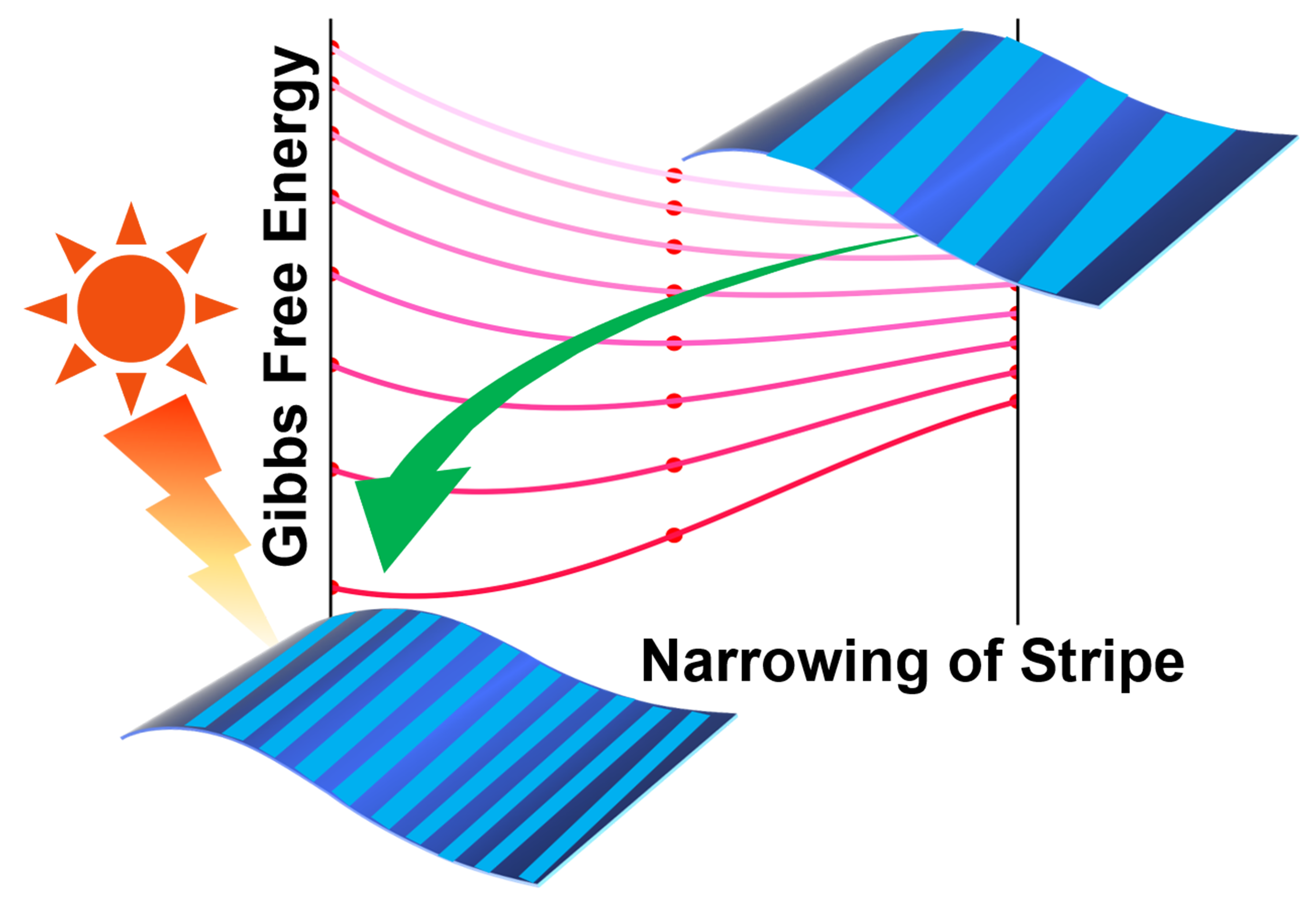}
\end{abstract}

\maketitle

\par
Ferroelectric (FE) materials, characterized by reversible spontaneous polarizations ($\vec{P}$) and multi-domains, have been widely studied and extensively explored for electronic devices, such as non-volatile information storage elements~\cite{RN6} and field-effect transistors~\cite{RN8} over the past few decades. The miniaturization requirements for large data storage density urge to reduce the ferroelectric materials into sub-nanometer (or atomic) scales. This has been, over a long time, believed to be unrealistic due to strong depolarization field effect~\cite{RN7}. Only until recently, scientists have theoretically predicted and experimentally demonstrated that low-dimensional (such as two-dimensional, 2D) intrinsic ferroelectric materials with nonvanishing and reversible dipole moments could exist in nature, e.g., group-IV monochalcogenide monolayers~\cite{RN2,RN5,RN11,RN31}, $\rm CuInP_{2}S_{6}$~\cite{RN10,RN13,RN14,RN4}, $\alpha-\rm In_{2}Se_{3}$~\cite{RN3,RN18,RN26,RN33}, $T'$-transition-metal dichalcogenide monolayers\cite{RN23,RN24}, and 2D perovskites~\cite{RN12,RN30,RN1,RN9}, to name a few.

\par
In addition to FEs, antiferroelectric (AFE) materials have also been attracting tremendous attention. Unlike FEs, the AFEs are immune to small external bias and exhibit a double electric-polarization $P-E$ hysteresis loop, which guarantees its high energy storage efficiency and large saturation polarization. Hence, AFE materials are also found to be promising for capacitors~\cite{RN19,RN20}, memristors\cite{RN21,RN22}, and piezoelectric\cite{RN4} applications. Compared with traditional bulk systems, 2D AFE systems are scarce, and more investigations remain largely under-explored. Recently, Xu \textit{et al.}~\cite{RN25} performed a seminal theoretically and experimentally combined work and disclosed a promising 2D AFE phase, $\beta'-\rm In_{2}Se_{3}$. Among the multiple phases in the polymorphic $\rm In_2Se_3$, the $\alpha$ and $\beta'$ phases could form 2D van der Waals (vdW) layers under room temperature ($\beta$ phase appears under high temperature~\cite{RN37}). These vdW layers consist of five atomic quintuple layers (Se-In-Se-In-Se), and they both belong to hexagonal lattice. Their difference lies at the atomically stacking patterns: The $\alpha$ phase forms A-B-B-C-A stacking sequence, while the $\beta$ (and $\beta'$) phase shows a merely A-B-C-A-B (fcc) order. Note that previous works have demonstrated that the precise fcc stacking sequence is dynamically unstable, and a shuffle of Se atom in the central atomic Se layer along $\langle 11\bar{2}0\rangle$ could remove its soft phonon mode~\cite{RN3,RN25}. In this regard, an in-plane FE order appears (denoted as $\rm FE\beta$ in this work). According to recent experimental observations~\cite{RN25,RN15}, this FE structure easily subjects to a phase transition and forms a $180^{\circ}$-AFE pattern (denoted as $\beta'$ here) with periodic antiparallel nanostripes, each of which contains a few $\rm FE\beta$ unit cells. Unlike in conventional 3D bulk (anti-)ferroelectric materials that consist micro-sized multi-domains along various directions, the AFE nanostripes in $\beta'-\rm In_2Se_3$ are exotic as they are in-plane dominant and in the atomic scale order. Hence, it could yield ultrahigh density information storage and facile manipulations in a subtle precision, meeting the requirements for sub-nanometer field-effect transistors and information storage devices in the near future. Yet, the modulation of the nanostripe patterns requires further evaluation and careful investigations. Due to the zero net electric polarization nature in the AFEs, traditional electric bias strategy is not straightforwardly appropriate.
\par
In this Letter, we propose that light-induced phase transition could occur in such AFE nanotripes, and the FE nanostripe width can be effectively manipulated. We illustrate this by performing first-principles density functional theory (DFT) calculations combined with optomechanical thermodynamic theory. Our results suggest that the nanostripe width of $\beta'-\rm In_2Se_3$ quintuple layer (QL) well-controls their relative stability. The optimal nanostripe width contains three unit cells at low temperature. Wider and narrower nanostripe AFE structures are slightly higher in total energy. Through a simple model, we propose how the AFE nanostripe interactions and the formation of domain walls control their internal energy. By systematically evaluating and comparing the electronic structures and electron contributed optical responses of these AFE systems, we reveal obvious optical contrasts in the near-infrared optical wavelength regime. Hence, under light illumination, we suggest that different optomechanical effect would occur, and one can realize a series of light-induced phase transitions to renormalize the FE nanostripe widths in the $\beta'-\rm In_2Se_3$. This suggests a noncontacting and noninvasive phase transition strategy which, to the best of our knowledge, have not been disclosed in AFE materials. In addition, we predict that such nanostripe variations can be detected by the AFE nanostripe dependent bulk photovoltaic (BPV) effect under above bandgap light illumination. Through evaluating the BPV photocurrent for each nanostripe, we show that opposite flowing photocurrents exist along the nanostripe direction.
\par
We perform DFT~\cite{M1,M2} calculations within the Vienna \textit{ab initio} simulation package (VASP)~\cite{M3,M4}. The exchange-correlation term in the Kohn-Sham equation is treated by the generalized gradient approximation (GGA) in the Perdew-Burke-Ernzerhof (PBE) form~\cite{M5}. Projector augmented-wave (PAW) method~\cite{M6} and a planewave basis set are used to describe the core and valence electrons, respectively, which yield a computational accuracy comparable to all-electron calculations. The kinetic energy cutoff is set to be $500\,\rm eV$, and the $\bm k$-mesh separation in the first Brillouin zone (BZ) is set to be less than $0.02\,\rm \AA^{-1}$ for each simulation supercell. The convergence criteria for the total electronic energy and Hellmann-Feynman force components are set as $1\times 10^{-8}\,\rm eV$ and $1\times 10^{-2}\,\rm eV/ \AA$, respectively. In order to eliminate the artificial interactions between periodic image layers, we add a vacuum space of $20\,\AA$ along the out-of-plane $z$ direction. Dipole-dipole interaction corrections have been applied~\cite{M14,M15}, hence the total energy and electronic structure quickly converge as the vacuum space thickness. To evaluate the BPV current, we fit the DFT calculated Hamiltonian using the maximally localized Wannier functions as implemented in the Wannier90 code~\cite{M7,M8}.
\par
According to our previous work~\cite{RN16,RN17}, when the photon energy is below the semiconductor bandgap, the optomechanical effects can be described by an extended thermodynamic theory that is based on a nonlinear optical process. It could trigger shape deformations and phase transitions in the reaction space, akin to the optical tweezer technology that moves atoms and molecules in the real space. The complex alternating electric field $\mathcal{\vec{E}}(\omega,t)=\vec{E}e^{-\mathrm{i}\omega t}$ of light applies work done per volume in the form of $du=\langle\vec{\mathcal{E}}(\omega,t)\cdot d\vec{\mathcal{D}}^{*}(\omega,t)\rangle$, where $\mathrm{i}=\sqrt{-1}$ is the imaginary unit, $\omega$ is the photon angular frequency, and the thermodynamic time-average $\langle\cdot\rangle$ is evaluated. Here, the $\vec{\mathcal{D}}(\omega,t)$ is electric displacement vector that is induced by $\vec{\mathcal{E}}(\omega,t)$, i.e., $\vec{\mathcal{D}}(\omega,t)=\varepsilon_{0}\tensor{\varepsilon}(\omega)\cdot\vec{\mathcal{E}}(\omega,t)$ ($\varepsilon_0$ is vacuum permittivity and $\tensor{\varepsilon}(\omega)$, being a second order tensor, is the frequency dependent dielectric function). As for the in-plane component in 2D systems where the electric field is used as natural variable, one evaluates the phase stability by its light-induced Gibbs free energy (GFE) density~\cite{RN38}
\begin{equation}\label{eq:GFE}
g=g_{0}-\frac{1}{4}\varepsilon_{0}\varepsilon_{ii}'(\omega)E_{i}^{2} \quad (i=x,y).
\end{equation}
Here $g_0$ is the GFE density before light is irradiated. Linearly polarized light (LPL) is assumed with its polarization along the $i$-direction. In this regard, the change of GFE density is scaled by the real part of the dielectric function component $\varepsilon_{ii}'(\omega)$ and the square of the applied alternating electric field magnitude $E_{i}$. According to random phase approximation, we calculate the frequency dependent dielectric function, which, in the framework of independent particle approximation~\cite{M9}, can be written as
\begin{equation}\label{eq:eps}
    \begin{split}
        \varepsilon_{ij}&=\delta_{ij}-\frac{e^2}{\varepsilon_0}\int_{BZ}\frac{d^3\bm k}{(2\pi)^3}\sum_{n,m}(f_{n\bm k}-f_{m\bm k})\\
        &\times\frac{\langle u_{n\bm k}|\nabla_{k_{i}}|u_{m\bm k}\rangle\langle u_{m\bm k}|\nabla_{k_{j}}|u_{n\bm k}\rangle}{\hbar(\omega_{m\bm k}-\omega_{n\bm k}-\omega-\mathrm{i}/\tau)},
    \end{split}
\end{equation}
where $\delta_{ij}$ is Kronecker delta symbol. $f_{n\bm k}$, $\hbar\omega_{n\bm k}$, and $|u_{n\bm k}\rangle$ refer to Fermi-Dirac distribution, eigenenergy, and the cell periodic wavefunction of band $n$ at $\bm k$. In order to phenomenologically incorporate the scatterings during electron motion, we adopt finite lifetime of $\tau=0.03\,\rm ps$ in the calculation. Note that according to Kramers-Kronig relation, the exact value of such lifetime only affects the resonant absorption frequency region. When the photon energy is much lower than the bandgap, the real part value of the dielectric function is marginally affected. Note that the denominator in this expression actually corresponds to the joint density of states, which indicates that larger bandgap would give a smaller dielectric function below bandgap. On the other hand, the numerator $\mathcal{M}_{ij}=\langle u_{n\bm k}|\nabla_{k_{i}}|u_{m\bm k}\rangle\langle u_{m\bm k}|\nabla_{k_{j}}|u_{n\bm k}\rangle$, usually referred to as interband transition dipole moment, also controls the optical selection between the valence and conduction bands. The synergistic effects between them determines the real part of dielectric function (or electronic contributed dielectric constants) below bandgap. The integral in Eq.~(\ref{eq:eps}) is performed in the 3D first BZ. For the 2D materials in a 3D simulation supercell, the vacuum contribution can be eliminated by multiplying a factor of $\frac{L_c}{d_{\rm eff}}$ where $L_c$ is the supercell lattice constant along $z$ and $d_{\rm eff}$ is the effective thickness of the QL (taking to be $10.5\,\AA$ here, measured from its corresponding vdW bulk structure). This re-scaling method has been widely adopted in previous works~\cite{RN16,RN17,M10}. Note that when we compare the relative stability (GFE), we will multiply effective volume to Eq.~\ref{eq:GFE}. Hence, the specific choice of $d_{\rm eff}$ value is cancelled which does not affect the thermodynamic evaluation.
\par
The imaginary part of dielectric function [$\varepsilon''(\omega)=\operatorname{Im}\varepsilon(\omega)$] determines the absorbance spectrum of 2D materials as
\begin{equation}
    A_{ii}(\omega)=1-e^{-\varepsilon_{ii}''(\omega)\frac{\omega L_c}{c_0}}\simeq\varepsilon_{ii}''(\omega)\frac{\omega L_c}{c_0}.
\end{equation}
Here $c_0$ is the speed of light in vacuum. One notes that the $A_{ii}(\omega)$ is independent to the vacuum space thickness in the simulation supercell.
\par
The BPV effect is a nonlinear optical process that exists in centrosymmetric ($\mathcal{P}$) broken systems~\cite{M11}. It arises from anharmonic motion of electrons and holes under above bandgap light excitation, which generates a static current with density of
\begin{equation}\label{eq:bpv}
\mathcal{J}^{i}=\sigma^{i}_{jk}(0;\omega,-\omega )E_j(\omega)E_k(-\omega),
\end{equation}
where $\sigma_{jk}^i(0;\omega,-\omega)$ is BPV photoconductivity. Under LPL irradiation ($j=k$), the dominant BPV effect is shift current generation (for nonmagnetic systems). According to our previous work~\cite{M12}, we can calculate the nanostripe-dependent BPV photoconductivity via
\begin{equation}\label{eq:sc}
    \begin{split}
        \sigma_{jj}^{i,D}&(0;\omega,-\omega)=\frac{e^{3}}{2\omega^{2}\hbar^{2}}\int_{BZ}\frac{d^{3}\bm k}{(2\pi)^{3}}\mathrm{Re}\sum_{mnl}(f_{l}-f_{m}) \\
        &\times\frac{v_{lm}^{j}}{\omega_{ml}-\omega+\mathrm{i}/\tau}\left[\frac{\tilde{v}_{mn}^{i,D}v_{nl}^{j}}{\omega_{mn}+\mathrm{i}/\tau}-\frac{v_{mn}^{j}\tilde{v}_{nl}^{i,D}}{\omega_{nl}+\mathrm{i}/\tau}\right].
    \end{split}
\end{equation}
All the quantities are $\bm k$-dependent, which are omitted for clarity reason. The velocity operator component is $v_{lm}^j=\langle u_{l\bm k}|\hat{v}^j|u_{m\bm k}\rangle=\frac{1}{\hbar}\langle u_{l\bm k}|\frac{\partial H}{\partial k_j}|u_{m\bm k}\rangle$, and $\omega_{lm}=\omega_{l\bm k}-\omega_{m\bm k}$ denotes the eigenenergy difference between band $l$ and $m$. For the AFE material that is $\mathcal{P}$-symmetric, the total $\sigma_{jj}^i=0$. In order to calculate the nanostripe-specific photocurrent, we use projected velocity operator for the $i$-propagating operator, namely, $\tilde{v}_{mn}^{i,D}=\langle u_{m\bm k}|\hat{P}_D\hat{v}^i|u_{n\bm k}\rangle=\sum_{\alpha\in D}\langle u_{m\bm k}|\psi_{\alpha}\rangle\langle\psi_{\alpha}|\hat{v}^i|u_{n\bm k}\rangle$, where $|\psi_{\alpha}\rangle$ denotes the Wannier function localized at the nanostripe-$D$.
\begin{figure*}[t]
    \centering 
    \includegraphics[width=0.9\textwidth]{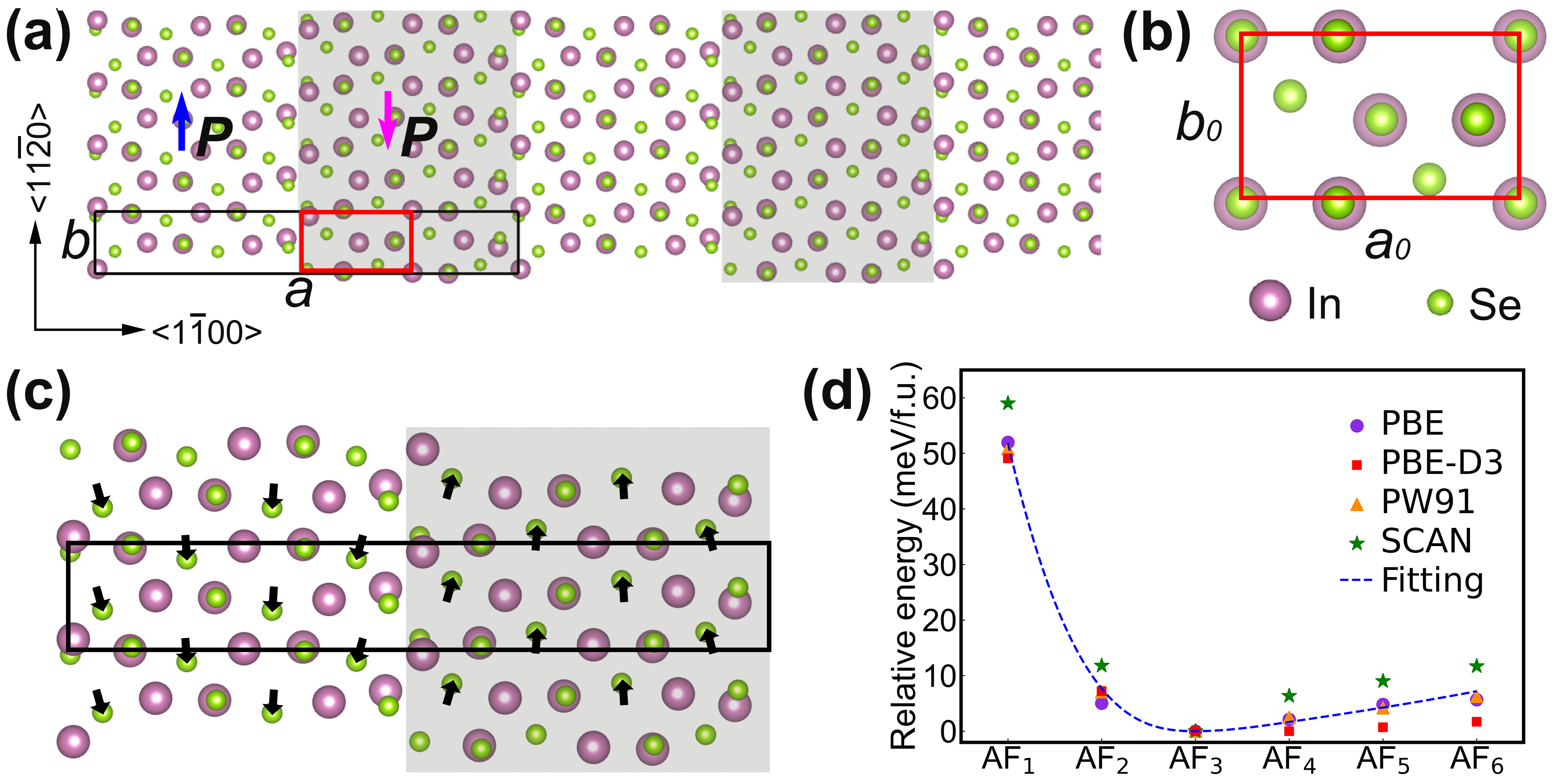}
    \caption{(a) Atomic structure of $\beta'-\rm In_2Se_3$ with AFE nanostripes. The black rectangle represents the simulation supercell which contains four parental $\rm FE\beta$ unit cells (indicated by a red rectangle). The spontaneous polarization is represented by the up and down arrows. Shaded area denotes the nanostripe with polarization along $-y$, for clarity reason. (b) Top view of the parental $\rm FE\beta$ unit cell. (c) Displacement of the central $\rm Se$ atoms relative to the high symmetric fcc structure (Figure S1a-b), denoted by black arrows. In (d) we plot the relative energy per f.u. of $\mathrm{AF}_n$ with $n=1,\,2,\,\dots\,6$, calculated by PBE (purple circles), PBE with D3-vdW correction (red square), PW91 (orange triangles), and SCAN (green stars) functionals. The fitting results of Eq.~(\ref{eq:energy}) for PBE results are plotted as blue dashed curve.}
    \label{fig:struct}
\end{figure*}
\par
In Figure \ref{fig:struct} we plot a typical atomic structure of AFE $\beta'-\rm In_2Se_3$. Our geometric relaxation yields similar results as the dark-field scanning transmission electron microscopy images reported previously~\cite{RN25,RN15}. In this case, the simulation supercell (the black rectangle in Figure \ref{fig:struct}a) contains two antiparallel nanostripes, polarized along $\pm y$ ($\langle 11\bar{2}0\rangle$). Here, each nanostripe is composed by two unit cells of the parental $\rm FE\beta$ structure (Figure \ref{fig:struct}b) along the armchair ($x$ or $\langle 1\bar{1}00\rangle$) direction, and the simulation supercell contains four $\rm FE\beta$ unit cells. We thus denote this structure as $\rm AF_4$. The optimized supercell lattice constants are $a=27.92\,\AA$ and $b=4.11\,\AA$. Each nanostripe width is $\sim$$1.4\,\rm nm$, well consistent with the experimental observation~\cite{RN25,RN36}. Note that the relaxed lattice constants of the parental $\rm FE\beta$ unit cell are $a_0=6.88\,\AA$ and $b_0=4.12\,\AA$. Hence, after forming AFE nanostripes, the in-plane strains exerted along the $x$ and $y$ are $1.5\%$ (tensile) and $-0.2\%$ (compressive), respectively. If the layer is freely suspended with slight pre-existing slacks, these small strains can be easily released to the vacuum space in the $z$ direction. This is different from the strain accumulations in traditional 3D bulk materials, which adds the reversible advantageous for 2D vdW materials. Hence, such strain effects can be omitted in the subsequent phase transition discussions.
\par
Compared with the high symmetric fcc phase [$P\bar{3}m1$ as in Figure S1a and S1b, Supporting Information (SI)], both In and Se atoms slightly shuffle in the AFE structure. Our calculations show that the displacement on each atom ($\delta\bm r_i$ with $i$ denoting the atom index) is mainly in the $xy$-plane ($\delta z_i\simeq 0$). We use small arrows to mark these displacements on the central Se atoms in Figure \ref{fig:struct}c, while those for other atomic layers are smaller and not shown here. One clearly observes that these displacements are mainly along the $y$ direction, and very small $x$ displacements appear on the Se atoms that are close to the AFE domain boundary. This reproduces the $180^{\circ}$ domain wall that reported previously~\cite{RN25}. We roughly estimate the polarization for each nanostripe-$D$ via $P_y^D=\frac{e}{\Omega}\sum_{\kappa\in D}Z_{\kappa,yy}^*\delta y$ where $\Omega$ is the cell volume and $Z_{\kappa,yy}^*$ is the Born effective charge component of ion-$\kappa$ (of each nanostripe). The calculated Born effective charge is listed in Table S1 (SI). As expected, the $P_x^D$ is found to be zero. The $y$ component polarization for each AFE domain is evaluated to be $P_y^D=2.3\times 10^{-10}\,\rm C/m$, comparable with previous results~\cite{RN27} and other 2D in-plain FE materials~\cite{RN32,RN34}.
\par
As one can see in the experimental observation, AFE nanostripes with different widths could emerge~\cite{RN25}. In order to evaluate and compare their relative stability, we construct and relax AFEs with different nanostripe domain widths, namely, $\mathrm{AF}_n\,(n=1-6)$. Their atomic structures are plotted in Figure S2. After careful geometric optimizations, we find that they all could sustain without any further spontaneous structural distortions. To confirm their spontaneous polarization, we estimate the polarization of each nanostripe, as listed in Table~\ref{tab:basic}. We plot their relative energies using pink scatters in Figure \ref{fig:struct}d. One sees that the $\rm AF_1$ is energetically higher than the other phases ($n=2-6$), which are within $10\,\rm meV$ per formula unit (f.u.). This is due to the high concentration of FE domain walls in $\rm AF_1$, and their interactions cause significant geometric distortions, as compared to the other structures ($n=2-6$). The small energy differences for larger $n$ systems indicate that phase transition may occur under weak stimulation. Among these structures, the $\rm AF_3$ takes the lowest calculated energy. In order to verify this, we optimize the geometries and calculate their total energies using other treatments for the exchange-correlation functional such as PW91~\cite{RN40} and SCAN (strongly constrained and appropriately normed)~\cite{RN39} functionals, and the PBE-D3 method~\cite{RN42} to incorporate the vdW correction. They give qualitatively consistent results as the PBE functional. On the contrary, one sees that the $\rm AF_4$ appears to be the most optimal pattern in experimental observations~\cite{RN25}, which is found to be slightly higher in energy according to our calculations. We attribute such discrepancy due to the entropy contributions since the DFT calculations are zero temperature results which do not include electronic and ionic entropy under finite temperature. Unfortunately, numerically computing ionic vibrational entropy in such large simulation supercells are extremely computational demanding and cannot be well-performed. Hence, our calculated total energy sequence is only valid under low temperature, while thermal effects under room temperature are omitted here. Nonetheless, as will be estimated in the following, the light-induced optomechanical energy dominates in the GFE variation over the other contributions. Thus, our main conclusion remains unchanged, regardless the inclusion of entropy or not. \par
In order to understand the total energy variation (Figure \ref{fig:struct}d), we split the total energy contributions into two sources, namely, the polarization exchange interaction between two neighboring nanostripes and the concentration of domain wall formation (or domain wall interaction). We express the total energy density as
\begin{equation}\label{eq:energy}
    E(\mathrm{AF}_n)=J\sum_{\langle i,j\rangle}P_y^iP_y^j+\frac{M}{\lambda}.
\end{equation}
Here $P_y^i$ (and $P_y^j$) denotes each distorted central Se contributed $y$-polarization (so that it spans $a_0/2$ along $x$), and the summation runs over all nearest neighbor. $J$ represents the 1D polarization exchange interaction parameter (similar as in the magnetic Ising model). $\lambda$ denotes the nanostripe length, and their values for each state are listed in Table~\ref{tab:basic}. The second term measures the formation of domain wall density contributions. Our fitting (for PBE) yields $J=22.3\,\mathrm{mV\cdot \AA^{2}\cdot C^{-1}\cdot f.u.^{-1}}$ and $M=0.5\,\rm{eV}\cdot \AA\cdot f.u.^{-1}$. The results are plotted as blue curve in Figure \ref{fig:struct}d. The positive $J$ value implies that antiparallel nanostripes (AFE aligned) is preferred, while the positive $M$ term prevents very dense domain wall formation. Therefore, these competitive interactions lead to an optimal AFE nanostripe width, namely, $\rm AF_3$ in our study. We also find that these two parameters, $J$ and $M$, are independent of optomechanically induced Gibbs free energies.
\begin{figure*}[htb]
    \centering 
    \includegraphics[width=0.9\textwidth]{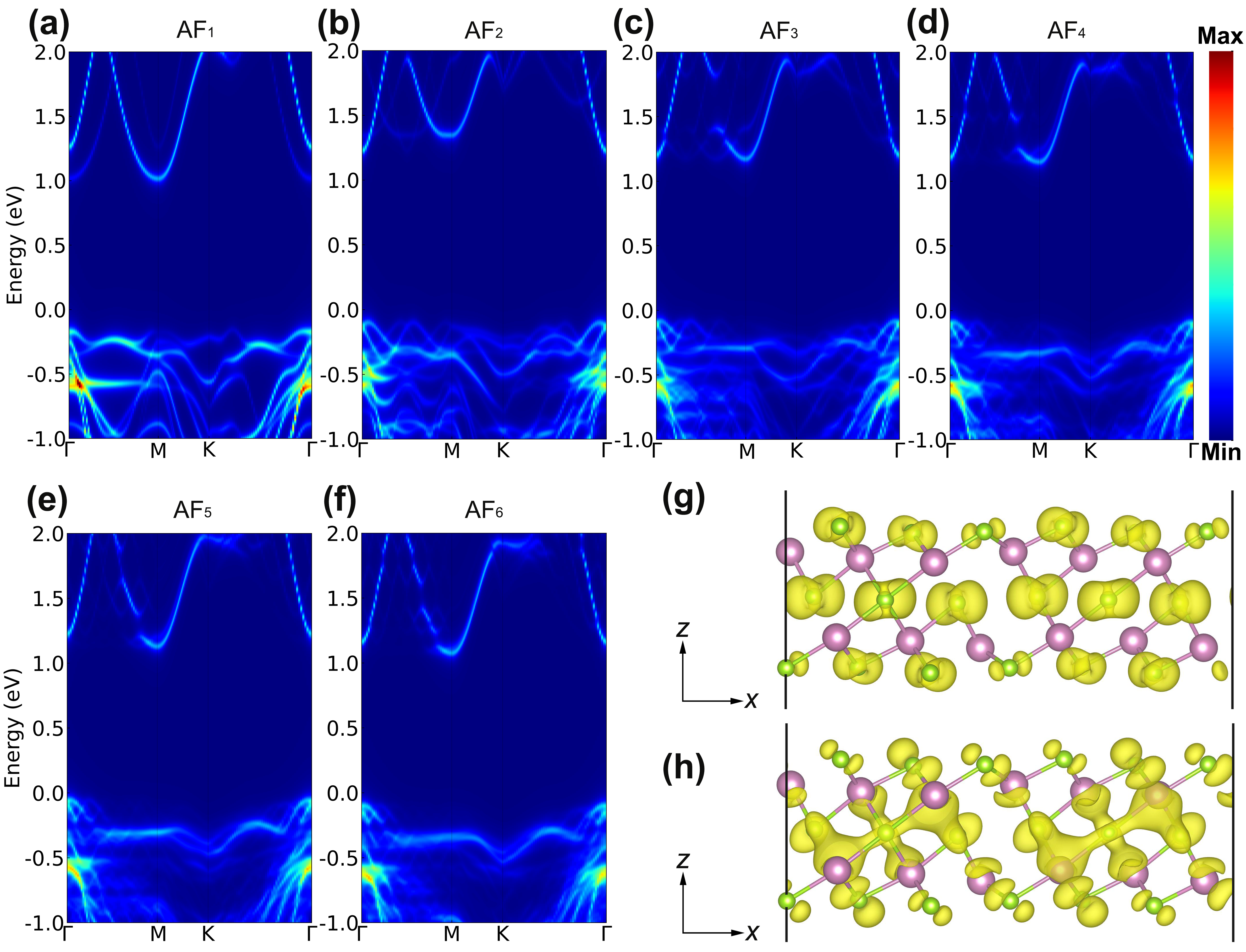}
    \caption{(a)-(f) Unfolded electronic band dispersion of $\mathrm{AF}_n\,(n=1-6)$ along the high symmetric $\bm k$-path of the first BZ of the parental fcc primitive cell ($P\bar{3}m1$ phase). The direct coordinates are $\Gamma=(0,0,0)$, $M=(1/2,0,0)$, and $K=(1/3,1/3,0)$. The band-decomposed charge densities of (g) VBM and (f) CBM in $\rm AF_3$ are also plotted.}
    \label{fig:band}
\end{figure*}

\begin{table}[htb]
    \caption{Basic physical quantities of $\mathrm{AF}_n$. Here $\lambda$, $P_y$, $E_g^{\rm ind}$, $E_g^{\rm d}$, and $\varepsilon'(\omega_0)$ denote the nanostripe width, polarization of an individual nanostripe, indirect and direct bandgaps, and the real part of dielectric function (at an incident energy of $\omega_0=0.8\,\rm eV$, averaged along the $x$ and $y$ directions), respectively.}
    \centering
    \begin{tabular}{c c c c c c c}
        \hline\hline
        $n$ & $1$ & $2$ & $3$ & $4$ & $5$ & $6$ \\
        \hline
        $\lambda\,(\AA)$ &$3.49$&$7.02$&$10.50$&$13.96$&$17.38$&$20.84$ \\
        $P_{y}\,(10^{-10}\rm{C/m})$ &$3.16$&$2.36$&$2.39$&$2.30$&$2.16$&$2.05$\\
        $E_g^{\rm ind}\,(\rm eV)$ &$1.07$&$1.33$&$1.26$&$1.24$&$1.19$&$1.18$\\
        $E_g^{\rm d}\,(\rm eV)$ &$1.12$&$1.35$&$1.26$&$1.25$&$1.19$&$1.18$\\
        $\varepsilon'$ &$12.76$&$11.81$&$11.17$&$11.36$&$11.43$&$10.98$\\
        \hline\hline
    \end{tabular}
    \label{tab:basic}
\end{table}
\par
We next explore their electronic band dispersion. Since each simulation supercell contains multiple unit cells and their corresponding BZs are different, we use the effective band structure method~\cite{M13} to unfold the supercell calculated band structures into the first BZ of $P\bar{3}m1$ primitive cell (Figure S1c in SI). During the unfolding process, the marginal strain effects are ignored. We plot the unfolded band dispersion for each AFE in Figure \ref{fig:band}a-\ref{fig:band}f. One sees that they are all semiconductors with bandgap values above $1\,\rm eV$. The valence band maximum (VBM) locate in the vicinity of the $\Gamma$ point in all cases. For the conduction band minimum (CBM), it locates at the $M$ point for AF$_{1,3-6}$, while it is at $\Gamma$ in AF$_2$. We tabulate their direct $E_g^{\rm d}$ and indirect $E_g^{\rm ind}$ bandgaps in Table~\ref{tab:basic}. One observes that these two values are similar in each case, indicating their quasi-direct bandgap semiconductor feature. In order to correct the well-known bandgap underestimation in PBE functional, we adopt the HSE06 hybrid functional~\cite{RN41} to calculate the band dispersions for $n=1-4$ (Figure S3 and Table S2). It can be seen that the bandgap variation trends from both methods are the same, even though the HSE06 gives larger bandgap values than PBE. In order to visualize the orbital contribution feature, we plot the band-decomposed charge densities for the $\rm AF_3$ as an example. Figures \ref{fig:band}g and \ref{fig:band}h illustrate that the VBM is mainly contributed from the Se atoms, while the CBM is delocalized around all Se and In atoms. This indicates real-space center mismatch in the valence and conduction bands. Hence, upon interband excitation, we may expect a static charge current generation. The change distribution of other structures show similar behavior and are plotted in Figure S4 (SI).
\begin{figure*}[tb]
    \includegraphics[width=0.95\textwidth]{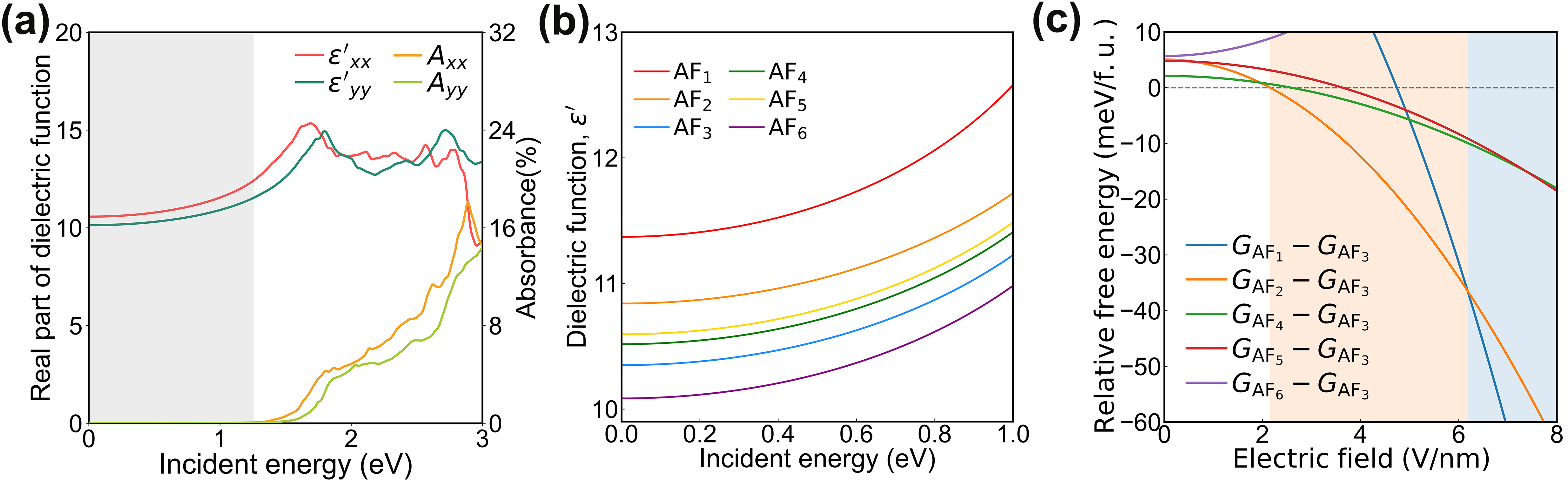}
    \caption{(a) Real part of frequency dependent dielectric function components and absorbance spectrum of $\rm AF_3$. The abscissor axis denotes the incident photon energy $\hbar\omega$. Grey shaded area denotes below bandgap energy. (b) Real part of averaged dielectric function $\varepsilon'(\omega)$ for $\mathrm{AF}_n\,n=1-6$ in the low frequency regime. (c) Relative GFE density variation under alternating electric field magnitude. The photon energy $\hbar\omega_0$ is taken to be $0.8\,\rm eV$. The white, orange, and blue background regions indicate that the $\rm AF_3$, $\rm AF_2$, and $\rm AF_1$ become the optimal structure, respectively.}
    \label{fig:opt}
\end{figure*}
\par
The electronic structure contrasts in the $\mathrm{AF}_n$ imply different optical responses. In order to explore this effect, we evaluate the ion-clamped dielectric function and the absorbance spectra. The results for $\mathrm{AF}_3$ is depicted in Figure \ref{fig:opt}a, and those for other structures are plotted in Figure S5 (SI). One sees that below the bandgap (shaded area), the direct light absorbance is exactly zero. Hence, in this frequency regime, the light-matter interaction is dominated by the real part of dielectric function $\varepsilon'(\omega)$. We observe slightly anisotropic optical response functions, owing to the contrast interband transition dipole moment $\mathcal{M}_{ii}=|\langle u_{c\bm k}|\nabla_{k_i}|u_{v\bm k}\rangle|^2$ in the $x$ and $y$ directions. Nonetheless, such anisotropy is marginal so that we will consider the in-plane averaged dielectric function $\varepsilon'(\omega)=[\varepsilon_{xx}^{'}(\omega)+\varepsilon_{yy}^{'}(\omega)]/2$ in the following discussion, which assumes that the optomechanical light is unpolarized and normal incident (propagating along $z$).
\par
In Figure \ref{fig:opt}b we compare the real part of in-plane averaged dielectric functions for different systems. One sees that the ${\varepsilon}'(\omega)$ for $\rm AF_1$ is larger than the other systems owing to its holding the smallest bandgap, as discussed previously. The $\rm AF_6$ takes the smallest $\varepsilon'(\omega)$ below the bandgap, indicating that under light irradiation, its GFE reduces slowliest among all systems. When we take an incident photon energy of $\hbar\omega_0=0.8\,\rm eV$ (or $1.55\,\mu\mathrm{m}$ in wavelength), their relative GFEs as functions of alternating electric field magnitude are plotted in Figure \ref{fig:opt}c. Note that at this incident energy, there is no direct light absorption, which corresponds to the imaginary part of dielectric function. Such direct light absorption would generate electron-hole pairs in the valence and conduction band, which usually subjects to a non-radiative recombination and produces unwanted waste heat. In our current approach, we use the real part of dielectric function, and keep its imaginary part to be zero (photon energy below bandgap and above phonon frequency regime). One can furthermore understand the mechanism by expanding the induced electric displacement under electric field according to the normal coordinate system, namely, $\vec{D}=\sum_{j,\bm q}(\tensor{\varepsilon}+\frac{\delta\tensor{\varepsilon}}{\delta{\bm X}_{j\bm q}}\cdot\delta{\bm X}_{j\bm q})\cdot\vec{E}$. Here, $\delta{\bm X}_{j{\bm q}}$ is normal coordinate displacement. $j$ and $\bm q$ are the phonon index and momentum, respectively. In this way, the light-induced free energy density variation (the second term in Eq. (\ref{eq:GFE}) is
\begin{equation}\label{eq:energy2}
    \delta g_{\vec{E}}=-\frac{1}{4}\sum_{j\bm q}\vec{E}\cdot(\tensor{\varepsilon}+\frac{\delta\tensor{\varepsilon}}{\delta{\bm X}_{j\bm q}}\cdot\delta{\bm X}_{j{\bm q}})\cdot\vec{E},
\end{equation}
Since the light wavelength is much longer than the unit cell scale, we approximate the momentum $\bm q$ to be nearly zero (${\bm q}\simeq 0$). The ``general force'' that applies onto the system is $\vec{f}_{j{\bm q}}=-\frac{\delta g_{\vec{E}}}{\delta{\bm X}_{j{\bm q}}}=\frac{1}{4}\vec{E}\cdot\frac{\delta{\tensor{\varepsilon}}}{\delta{\bm X}_{j{\bm q}}}\cdot\vec{E}$. This corresponds to a Raman-type displacement, which is an inelastic two-photon-one-phonon process, where the photon energy difference equals to the phonon energy. Therefore, the emitted photon is red-shifted compared with the incident light, and the number of photons is unchanged during the whole process (no direct absorption). The general force pushes the system to transit into a state that has a larger dielectric function value. This converts the light energy into structural mechanical energy, so that the phase transition is optomechanical. In an ideal situation, the light induced work done totally converts into phase transition, so that no waste heat is generated and the temperature increment is kept to be very low. Note that the phase transformation process would still produce temperature variation due to entropy contrast between two phases, but it should be much smaller as compared with the direct light absorption process. According to Figure \ref{fig:opt}c, one sees that at an intermediate electric field intensity (below $2.2\,\rm V/nm$ or $6.2\times 10^{11}\,\rm W/cm^2$), the $\rm AF_3$ has the lowest GFE. Above this magnitude, the $\rm AF_2$ becomes the most thermodynamically stable structure, indicating its emergence under intermediate light irradiation (shaded in orange). When the light intensity is above $5.1\times 10^{12}\,\rm W/cm^2$ (alternating electric field magnitude of $6.2\,\rm V/nm$), the $\rm AF_1$ appears as the most optimal structure (shaded in blue). Therefore, we elucidate that a near-infrared light irradiation could gradually reduce the AFE nanostripe widths. Such phase transition does not need long range atomic redistribution as in the sluggish crystalline-noncrystalline phase transformations, hence it belongs to diffusionless fast martensitic phase transition. We note that wider nanostripe width structures would not appear under light irradiation due to their relatively lower dielectric function values. We also perform nudged elastic band calculations and reveal that the energy barrier separating different AFE width phases is $\sim 0.15\,\rm eV$ per f.u. Note that this energy barrier is not sufficiently large, so that the $\rm AF_2$ may only survive under low temperature for a long time.

\begin{figure*}[tb]
    \centering 
    \includegraphics[width=0.95\textwidth]{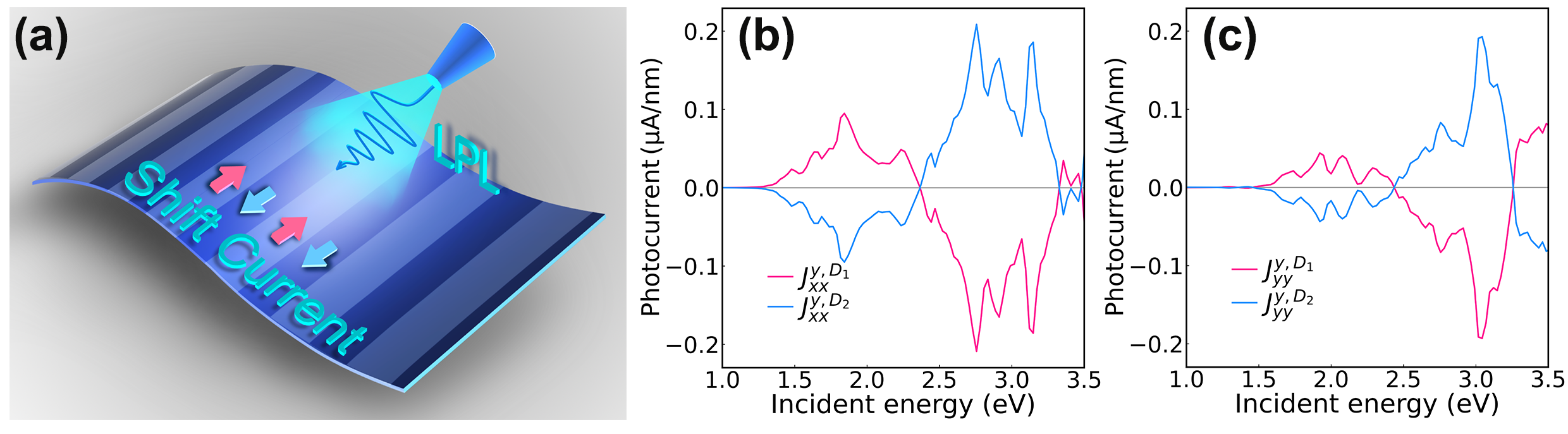}
    \caption{(a) Schematic plot of BPV current generation (under LPL) and detection. AFE nanostripe-specified shift current density for $\rm AF_3$ under (b) $x$-LPL and (c) $y$-LPL are also plotted. We assume that the above bandgap light intensity is $0.1\,\mathrm{V/nm}$. The abscissor axis is incident photon energy.}
    \label{fig:sc}
\end{figure*}
\par
In order to precisely detect such nanostripe width variation and explore its potential applications, we suggest to resort to nanostripe-dependent BPV effect, which could convert weak above bandgap light into static electric current (or low frequency, usually THz, emission). According to nonlinear optics theory, LPL could induce shift current generation in nonmagnetic materials, which arises from the real space wavefunction center mismatch in the valence and conduction bands. In this way, optical excitation between them results in the shift of such centers and generates electric current. Such a process also reflects the topological feature of electron wavefunctions. We propose that AFE nanostripe would produce opposite flowing shift current (Figure~\ref{fig:sc}a), hence one may detect nanostripe-specific short-circuit currents or open-circuit voltages. We take the LPL with a field intensity of $0.1\,\rm V/nm$ and calculate BPV photocurrent per stripe width. The results for $\mathrm{AF}_3$ is plotted in Figures~\ref{fig:sc}b (under $x$-polarized LPL) and \ref{fig:sc}c (under $y$-polarized LPL). First of all, we find that the total BPV current (summed over two antiparallel nanostripes) is always zero, consistent with the fact that the whole AFE $\beta'-\rm In_2Se_3$ is centrosymmetric. However, inside each nanostripe, one observes that the $\mathcal{J}_{ii}^{y,D_1}=-\mathcal{J}_{ii}^{y,D_2}\neq 0$ ($i=x, y$). Note that under $x$ or $y$ polarized light irradiation, the current only flows along $y$ (or $\langle 11\bar{2}0\rangle$, parallel to nanostripe), while no current occurs along $x$ ($\langle 1\bar{1}00\rangle$, normal to nanostripe direction). This can be ascribed by the mirror reflection (normal to $x$) symmetry of the whole system. The magnitude of the 1D current density in each nanostripe is on the order of $0.1\,\mu\mathrm{A/nm}$, which is sufficiently large for experimental detection. Similar photocurrents on the same magnitude order are also calculated in $\rm AF_1$ and $\rm AF_2$ (Figure S6 in SI). This nanostripe-specific photocurrent is similar as in the recently disclosed concept of hidden spin polarization~\cite{RN28}, hidden Berry curvautre~\cite{RN29}, and hidden Hall current~\cite{RN35}. In all these cases, the total response functions of the system are zero, subject to specific (usually inversion or time-reversal) symmetries. But once we project them into a partial sector, they would show finite and observable signals. The fundamental mechanism is that the local symmetry of each nanostripe is lower than that of the whole system, and the removal of mirror reflection along $x$ yields finite and observable BPV currents. Here, we show that the shift current generation in 2D AFE system also belongs to such paradigm.
\par
In conclusion, we scrutinize the light-induced structural transformation of $\beta'-\rm In_2Se_3$ nanostripes. According to the optomechanical theory and first-principles calculations, we suggest that one can use below bandgap light to effectively narrow the AFE nanostripe widths. By calculating the nanostripe-dependent shift current generations, we reveal hidden bulk photovoltaic shift current generations. Structural transformation of 2D ferroic materials is of great importance for fabricating advanced microelectronic devices. Our work provides a noncontacting and noninvasive strategy for manipulating AFE domains in atomic precision and would be helpful for designing nanoelectronic devices with large information storage density.

\section{Acknowledgments}
This work is supported by the National Natural Science Foundation of China (NSFC) under Grant Nos. 21903063 and 11974270. The computational resources provided by HPC platform of Xi'an Jiaotong University are also acknowledged.

\section{Supporting Information}
\textbf{Supporting Information Available:} Born effective charge, structure of fcc monolayer, supercells of AF$_n$, band structure calculated by HSE06 hybrid functional, band-decomposed charge densities, dielectric function, and shift current photoconductivity of AF$_1$ and AF$_2$.

\bibliography{Mainbib}
\end{document}